\documentclass[%
 preprint,
 amsmath,amssymb,
 aip,
]{revtex4-1}

\usepackage{graphicx}
\usepackage{dcolumn}
\usepackage{bm}
\usepackage{siunitx}

\begin{document}


\title{Collective Strong Coupling with Homogeneous Rabi Frequencies using a 3D Lumped Element Microwave Resonator}

\author{Andreas Angerer}
\email{andreas.angerer@tuwien.ac.at}
\affiliation{Vienna Center for Quantum Science and Technology, Atominstitut, TU Wien, Stadionallee 2, 1020 Vienna, Austria}

\author{Thomas Astner}
\affiliation{Vienna Center for Quantum Science and Technology, Atominstitut, TU Wien, Stadionallee 2, 1020 Vienna, Austria}

\author{Daniel Wirtitsch}
\affiliation{Vienna Center for Quantum Science and Technology, Atominstitut, TU Wien, Stadionallee 2, 1020 Vienna, Austria}

\author{Hitoshi Sumiya}
\affiliation{Sumitomo Electric Industries Ltd., Itami, 664-001, Japan}

\author{Shinobu Onoda}
\affiliation{Takasaki Advanced Radiation Research Institute, National Institutes for Quantum and Radiological Science and Technology, 1233 Watanuki, Takasaki, Gunma 370-1292, Japan
Japan}

\author{Junichi Isoya}
\affiliation{Research Centre for Knowledge Communities, University of Tsukuba, 1-2 Kasuga, Tsukuba, Ibaraki 305-8550,
Japan}

\author{Stefan Putz}

\affiliation{Vienna Center for Quantum Science and Technology, Atominstitut, TU Wien, Stadionallee 2, 1020 Vienna, Austria}
\affiliation{Department of Physics, Princeton University, Princeton, NJ 08544, USA}

\author{Johannes Majer}
\email{johannes.majer@tuwien.ac.at}
\affiliation{Vienna Center for Quantum Science and Technology, Atominstitut, TU Wien, Stadionallee 2, 1020 Vienna, Austria}

\date{\today}

\label{par:abstract}
\begin{abstract}

We design and implement 3D lumped element microwave cavities for the coherent and uniform coupling to electron spins hosted by nitrogen vacancy centers in diamond.  Our design spatially focuses the magnetic field to a small mode volume.  We achieve large homogeneous single spin coupling rates, with an enhancement of the single spin Rabi frequency of more than one order of magnitude compared to standard 3D cavities with a fundamental resonance at \SI{3}{GHz}. Finite element simulations confirm that the magnetic field component is homogeneous throughout the entire sample volume, with a RMS deviation of 1.54\%. With a sample containing $10^{17}$ nitrogen vacancy electron spins we achieve a collective  coupling strength of $\Omega=\SI{12}{MHz}$, a cooperativity factor $C=27$ and clearly enter the strong coupling regime. This allows to interface a macroscopic spin ensemble with microwave circuits, and the homogeneous Rabi frequency paves the way to manipulate the full ensemble population in a coherent way.
\end{abstract}

\maketitle


\section{\label{sec:level1}Introduction}

Many different approaches have been proposed for the realization of hybrid quantum systems \cite{xiang_hybrid_2013,wallquist_hybrid_2009,verdu_strong_2009,imamoglu_cavity_2009,kubo_hybrid_2011} ranging from atoms to solid state spins coupled to microwave cavities. Most of these systems are realized by using distributed transmission-line \cite{goppl_coplanar_2008} or zero-dimensional lumped element resonators \cite{hatridge_quantum_2013}, which allow small confined cavity mode volumes in the microwave domain. This gives rise to strong interactions of the electric or magnetic dipole moment of a two level system with the electromagnetic field of the cavity. In order to allow the coherent exchange of energy on the single photon level \cite{wallraff_strong_2004,Abe2011,Boero2013} the coupling strength of the single-mode cavity with the two level system has to exceed all dissipation rates in the system.

The interaction strength of the cavity field with a magnetic dipole is small compared to that of an electric dipole, therefore, the interaction has to be collectively enhanced. This can be realized by an ensemble of $N$ spins in the cavity mode whose collective coupling strength is increased by a factor \cite{thompson_observation_1992,brennecke_cavity_2007, colombe_strong_2007} $\sqrt{N}$. Strong coupling of such large spin ensembles to one-dimensional cavities has been successfully demonstrated \cite{Amsuss2011,Kubo2010,Schuster2010,zollitsch_high_2015,zhang_strongly_2014,huebl_high_2013}. However, a big disadvantage of these cavities arises when coupling to large ensembles of emitters, because in planar circuits the cavity magnetic field component has a large field gradient. A macroscopic sample placed on top of such a structure is emerged in a magnetic field which changes significantly over the samples spatial extend. This leads to a strongly inhomogeneous distribution of coupling strengths between individual spins and the cavity. The resulting inhomogeneous single spin Rabi frequencies make it impossible to coherently manipulate and control the whole spin ensemble.

A different approach is using 3D microwave cavities. \cite{paik_observation_2011, Probst2014,axline_coaxial_2016,reshitnyk_3d_2016} They offer the possibility of high Q values due to their closed structure\cite{walther_cavity_2006,creedon_3d_2016} and only negligible contributions to loss from surface two-level systems \cite{reagor_reaching_2013}, but suffer from their huge mode volume. This results in very weak single spin coupling strengths and requires large ensemble sizes or spin system with larger magnetic moments \cite{Probst2014}. Some of theses issues have been addressed already and have shown promising results. \cite{floch_towards_2016,creedon_strong_2015,minev_planar_2016}

We report the design of a 3D lumped element microwave cavity in order to ensure large and \textit{homogeneous} single spin Rabi frequencies. In combination with inhomogeneous broadening of the spin ensemble \cite{krimer_non-markovian_2014}, for which we can account for using spin-echo refocusing techniques, an inhomogeneous field distribution throughout the sample prohibits the uniform and coherent manipulation of a spin ensemble. Our design addresses this problem by making use of metallic structures to focus the AC magnetic field, such that the resulting field distribution is homogeneous throughout the mode volume while yielding large coupling rates at the same time. This allows us to couple to comparatively small spin ensembles in a uniform and coherent way.

In a numerical study, using COMSOL Multiphysics \copyright \,  RF-module,  we analyze the electromagnetic field distribution inside the cavity and optimize the coupling strength for an individual spin. The cavity design is fabricated and loaded with a high pressure high temperature (HPHT) diamond sample containing \SI{40}{ppm} of NV spins. By using our new design we reach the strong coupling limit of cavity QED and open the way for new cavity QED protocols.

\section{\label{sec:level1_3dcav}Bow-tie Cavities}

In cavity QED the interaction of light and matter is enhanced by creating boundary conditions for the electromagnetic field inside a spatially confined mode volume. In the optical domain this is usually achieved by using Fabry--P\'{e}rot like cavities which create standing waves in free space or inside a waveguide between two mirrors. This principle can be brought to the microwave regime in different ways. One of the most straight forward implementations is a rectangular wave guide cavity \cite{Pozar2011} where the eigenfrequency is determined by the box dimensions with respect to the wavelength of the cavity photons. This leads to large resonators with mode volumes of at least larger than $\lambda^3/8$ (with $\lambda$ the wavelength of the cavity photons). 

However, microwave electronics also uses on chip transmission line waveguides and the realization of one dimensional microwave cavities has proven as to be a powerful implementation from high efficiency single photon detection \cite{lita_counting_2008,kubo_electron_2012} to quantum information processing \cite{johnson_quantum_2010,barends_superconducting_2014}. These devices and their spatial extend is however always connected to the wavelength of the fundamental resonance, regardless if they are one or three dimensional distributed.

In contrast an electrical  $LC$ oscillator is created by discrete elements for capacitance $C$ and inductance $L$, for which the eigenfrequency is given by $\omega=\frac{1}{\sqrt{LC}}$. In radio and microwave technology the wavelength can become much larger then the actual size of the circuit which is known as lumped element circuit. This means that in a lumped element resonator (LER) \cite{hatridge_quantum_2013} the actual extend is not related to the wavelength which allows to modify the size and shape of the cavity mode volume. This principle can also be applied to three dimensional lumped structures \cite{creedon_strong_2015}. 

\begin{figure}[t]
\includegraphics[]{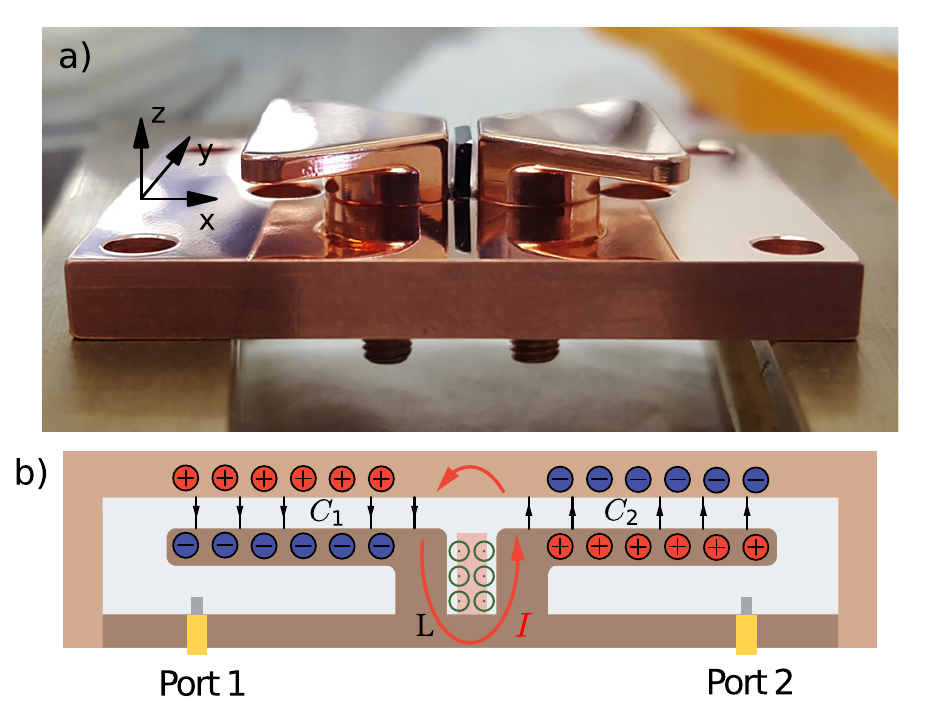}
\caption{a) Picture of our manufactured cavity with diamond sample glued to the cavity using vacuum grease. Note that for illustrative purposes the top lid that closes the structure and the sidewalls are not shown in this picture (Lighter brown sections in Fig. \ref{fig:cavity}b). b) Schematic cross section of the cavity perpendicular to the mode direction. Capacitors and inductance of the cavity are labeled as $C_1$, $C_2$ and $L$, with the electric and magnetic field generated by this design. The current that generates the focused magnetic field is drawn as red line. The diamond sample is drawn as shaded red rectangle.}
\label{fig:cavity}
\end{figure}

Here we demonstrate how this can be brought one step further by using our so-called "bow-tie" structure within a 3D cavity. These structures not only focus the magnetic field to a small mode volume, but also make it possible to couple strongly and homogeneous to a small spin ensemble (Fig. \ref{fig:cavity}a). The two bow-tie shaped elements are placed in a closed conducting box where the top surfaces form two large capacitors with the lid. The total capacitance of this cavity is then given by two capacitors in a series configuration reading
\begin{eqnarray}
C_{tot}^{-1} = C_1^{-1} + C_2^{-1} = 2C^{-1} = \frac{2}{\epsilon_0}\frac{d}{A},
\label{eq:cap}
\end{eqnarray}
where $A$ is the top area of one bow-tie, while $d$ is the separation between the bow-tie and the cover lid. 

The inductance $L$ is governed by the path length $l$ the current has to flow between the two capacitors in order to close the $LC$ circuit (Fig. \ref{fig:cavity}b).  Numerical simulations show that it can be approximated by the inductance of a flat wire $L_{tot}\propto l\left(\ln(l/w)+w/l\right)$ of width $w$ corresponding to the width of the bow-tie structures. The resulting eigenfrequency is then given by
\begin{eqnarray}
\omega_c = \frac{1}{\sqrt{L_{tot}C_{tot}}}.
\end{eqnarray}
Thus, the resonance frequency of our 3D LER is only governed by the geometric values of $l$, $w$, $A$ and $d$ and not the dimensions of the enclosing box. This allows to build resonators significantly smaller than the wavelength of our cavity photons. Employing this principle we design the cavity in a way that the distance between the bow-tie structures can be adjusted after manufacturing. With control over the inductance and capacitance we reach a wide range of tune-ability for the resonance frequency and flexibility concerning different sample sizes in the active region.

\begin{figure}[t]
	\includegraphics[]{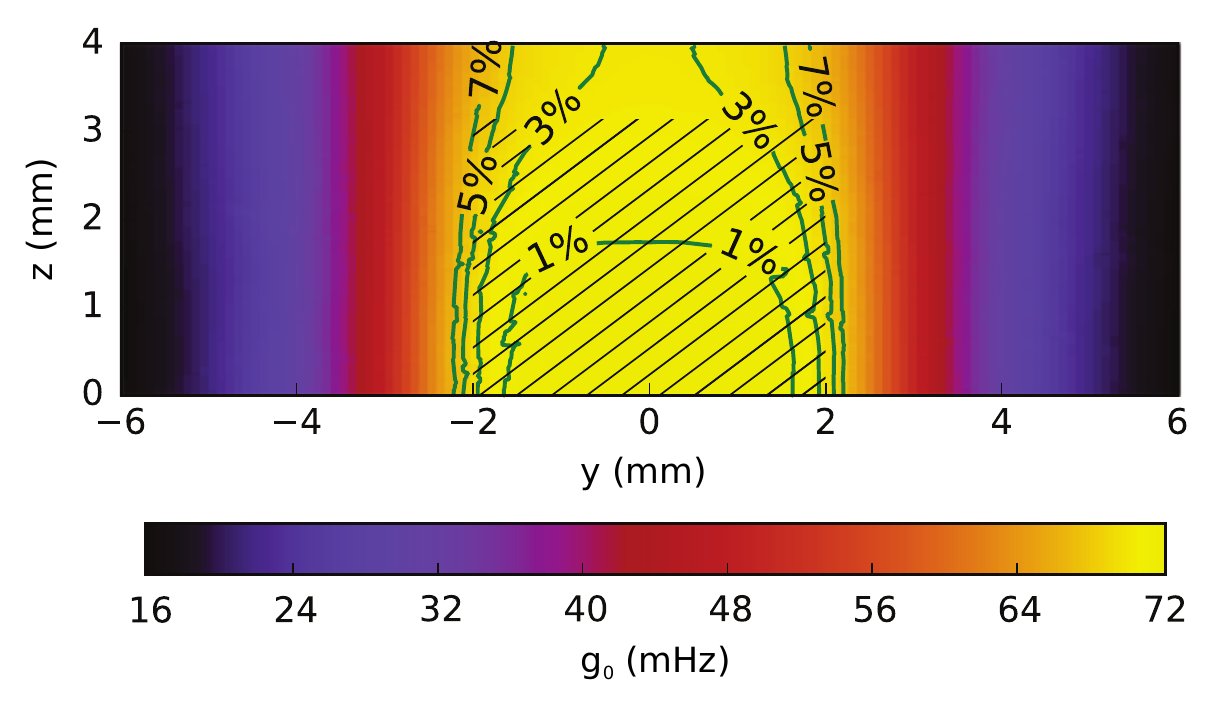}
	\linespread{1.}
	\caption{Illustration of the simulated magnetic field distribution. The plot shows a cross section of the mode volume parallel to the direction of the magnetic field mode (y-direction in Fig.\ref{fig:cavity}a). For the diamond sample used in this work (shown as dashed area) the deviation for the absolute value of the coupling strength is at most \SI{7}{\percent}, with a RMS deviation of 1.54\%. The green contour lines divide the area in sections wherein the magnetic field strength differs by a certain percentage.}
	\label{fig:homogeneity}
\end{figure}

The electromagnetic field inside the cavity is capacitively coupled via two coaxial ports, which are located below the bow-tie structures. In- and out-coupling capacitances can easily be adjusted by changing the length of these probes, controlling if the cavity is under or over coupled. The current is counter-propagating in the two bow-tie structures, therefore, the magnetic field between them is amplified and focused to a relatively small mode volume. This results in an enhancement of the magnetic dipole interaction of the cavity mode with our spin system. Moreover, from numeric simulations, we observe that the magnetic field is homogeneous within \SI{1.57}{\percent} in the region between the bow-tie structures (see Fig. \ref{fig:homogeneity}) when calculating the RMS deviation integrated over the sample volume. The improved homogeneity field distribution also ensures homogeneous single spin coupling rates $ g_0 $ over the entire sample volume.

From finite element simulations we can extract the electromagnetic field strength and distribution inside the cavity for a given input power. The resulting single spin cavity coupling rate is deduced from the total number of photons in the cavity mode for an specific input power since $n \approx E_{em}/\hbar\omega$. Using the simulated magnetic field strength $\vec{B}_{rf}$ normalized by the cavity photon number $n$ as $\vec{B}_{rf}^0=\vec{B}_{rf}/\sqrt{n}$, we can infer the coupling strength of a single spin magnetic moment with the vacuum cavity mode by
\begin{eqnarray}
\left|g_0\right| = \frac{\mu_B g}{2\hbar}\left|\vec{B}_{rf}^{0, \perp}\right|\left|\vec{S}\right|=\sqrt{\frac{2}{3}}\frac{\mu_B g}{2\hbar}\left|\vec{B}_{rf}^0\right|\left|\vec{S}\right|.
\end{eqnarray} 
Here $\vec{B}_{rf}^{0,\perp}$ stands for the perpendicular component of the vacuum magnetic field with respect to the spin principle axis, $\vec{B_{rf}^0}$ denotes the total vacuum magnetic field, $\mu_B$ the Bohr magneton, $g$ the gyromagnetic ratio that corresponds to the spin system used, $\left|g_0\right|$ the absolute value for the single spin coupling strength and the factor $\sqrt{2/3}$ accounts for the angle between cavity mode vector and spin principle axis. This yields values for the single spin coupling strength of $|g_0|\approx\SI{70}{mHz}$. These coupling rates are considerably higher compared to the coupling strengths of $\approx\SI{5}{mHz}$ achievable in a standard 3D rectangular waveguide cavity in the same frequency region.

\section{Measurement}\label{sec:measurement}

\begin{figure*}[ht]
\includegraphics[width=\linewidth]{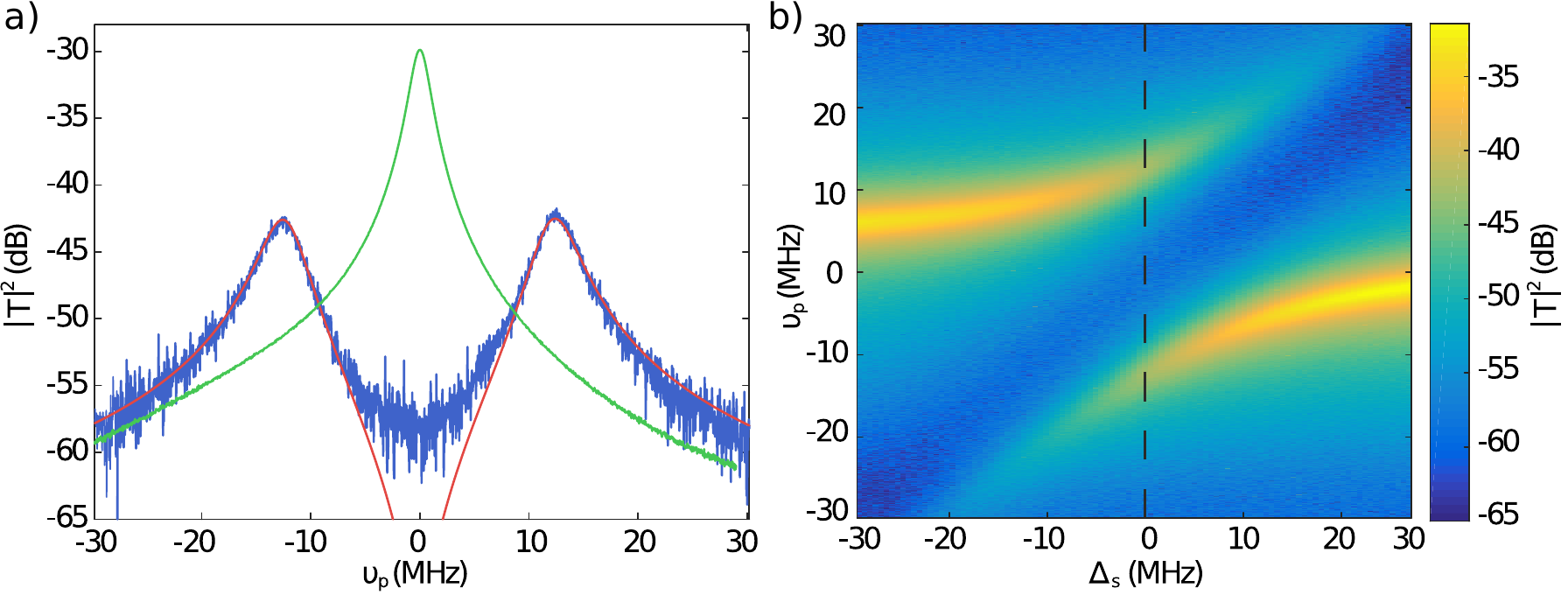}
\linespread{1.}
\caption{a) Transmission measurements for the cavity with the spin system far detuned (\textit{green curve}) and on resonance with an observed normal mode splitting observed for four NV sub-ensembles (\textit{blue curve}). Fitting the normal mode splitting (\textit{red curve}) with Eq. \ref{eq:trans} yields a collective coupling strength of $\Omega = \SI{12.46}{MHz}$. b) Cavity transmission spectroscopy as a function of $\nu_P$ versus  $\Delta_s$. Here $\nu_P$ stands for the detuning of the probe frequency relative to the spin frequency and $\Delta_s$ for the detuning of the cavity with respect to the spins. The dashed line denotes the detuning for which the normal mode splitting is measured}
\label{fig:transspec}
\end{figure*}

For our measurements we adjust the length of the coupling ports such that the cavity is effectively under-coupled and the total quality factor $Q$ is dominated by the internal cavity photon damping rate. These Cavity intrinsic losses are dominated by ohmic losses in the metallic structures, which are proportional to surface roughness \cite{proekt_investigation_2003} and skin-depth of the AC magnetic field. To allow external magnetic fields to penetrate the cavity, which is necessary for Zeeman tuning electron spins into resonance with the cavity, we fabricate it using oxygen free copper ($99.997\%$ purity). In order to further reduce ohmic losses from the metallic structures we perform a mechanical surface treatment and polish all surfaces to a roughness $<\SI{0.25}{\micro m}$. This yields, for a fundamental frequency of the unloaded cavity of \SI{2.775}{GHz} a quality factor of $ Q = 1920$ at $\SI{1}{\kelvin}$ (see Fig. \ref{fig:transspec}a).

Next we load the cavity with a diamond sample containing a large ensemble of negatively charged nitrogen-vacancy (NV) defect centers. NV centers are formed by a substitutional nitrogen atom and an adjacent vacancy in the diamond lattice. In the following we use a synthetic type-Ib high pressure high temperature (HPHT) diamond with an initial concentration of $\SI{100}{ppm}$ nitrogen. The sample is irradiated with \SI{2}{MeV} electrons at \SI{800}{\degreeCelsius} and annealed at
\SI{1000}{\degreeCelsius} multiple times with a total electron dose of  \SI{1.1e19}{cm^{-2}}. After electron irradiation and annealing this results in a NV concentration of approximately $\SI{40}{ppm}$. 

Four different sub-ensembles depending on the crystallographic location in a diamond cell are pointing in the $\langle 1,1,1 \rangle$ direction. Its electronic spin 1 ground state triplet can be described by a Hamiltonian of the following form
\begin{eqnarray}\label{eq:hamil}
H_{NV} = h D S_z^2 + g \mu_B \vec{B}_0\cdot \vec{S}.
\end{eqnarray} 
with a zero-field splitting of $D/h=\SI{2.87}{GHz}$ for the axial component along the NV-axes. The second term accounts for the interaction with an static external magnetic field ($\vec{B_0}$), which we use to lift the degeneracy of the $m_s = \pm 1$ manifold and tune the spin transitions into resonance with the cavity mode. 

The diamond sample is glued into the cavity using vacuum grease. The (1,0,0) face is placed parallel to the magnetic field mode direction ensuring the projection of the cavity mode onto the four possible NV axes to be equal. The diamond loaded cavity is mounted inside a dilution refrigerator operating at $\SI{25}{\milli \kelvin}$ in order to achieve thermal polarization rates of our spin ensemble in the ground state well above $99\%$. A 3D Helmholtz coil configuration provides arbitrary d.c. magnetic fields for Zeeman tuning our spin ensemble into resonance with the cavity later on.

We use a vector network analyzer to probe the transmission of the coupled cavity spin system in the stationary state. The diamond loaded cavity has a fundamental resonance frequency of $\nu=\SI{3.121}{GHz}$ with a $Q$ value of 1637 and cavity linewidth of $\kappa=\SI{1.91}{MHz}$ (HWHM), measured with all spins far detuned from the cavity resonance frequency. By applying a homogeneous external d.c. magnetic field on the whole ensemble in the [0,1,0] direction, we tune all four possible NV sub-ensembles into resonance with the cavity mode. The measured scattering parameter $|S_{21}|^2$ is compared to an analytical expression for the scattering parameter obtained from a Jaynes-Cummings Hamiltonian reading
\begin{eqnarray}\label{eq:trans}
|S_{21}|^2 = \left|\kappa\frac{\left(\omega-\omega_s-i\gamma^*\right)}{\left(\omega-\omega_c-i\kappa\right)\left(\omega-\omega_s-i\gamma^*\right)-\Omega^2}\right|^2,
\end{eqnarray}
with $\gamma^*$ the inhomogeneously broadened linewidth of our spin ensemble, $\omega$ the probe frequency and $\omega_c$ the cavity resonance frequency (see Fig. \ref{fig:transspec}). At the point where the central spin transition is in resonance with the cavity mode, we observe an avoided crossing which corresponds to a collective coupling strength of $\Omega=\SI{12.46}{MHz}$. This is in good agreement with our simulations, assuming a homogeneous single spin coupling strength of $\SI{70}{mHz}$ and a number of spins coupling to our cavity as $\approx 10^{17}$. The number of spins can be deduced from the sample volume of $\SI{4.2x3.4x0.92}{mm}$ and an approximate NV density of \SI{40}{ppm}. 

The collective coupling strength is large enough to satisfy the strong coupling condition $\Omega>\kappa, \gamma^*$ (HWHM, see Fig. \ref{fig:transspec}a) with an estimated inhomogeneously broadened spin linewidth of $\gamma^* \approx \SI{3}{MHz}$ (HWHM). Note that we are neglecting effects like the cavity protection effect \cite{putz_protecting_2014} and treat $\gamma^*$ to be a constant to the first order. This gives a cooperativity parameter of approximately $C=\Omega^2/(\kappa\gamma^*)\approx27$.

Our results show that we successfully enter the strong coupling regime of cavity QED using our 3D lumped element cavity design. The homogeneous single spin Rabi frequency allows coherent manipulation of the whole ensemble, taking into account the inhomogeneously broadened linewidth. Spin control protocols no longer suffer from excitations diffusing out of the active region via spin-spin interaction, since all spins are emerged in a cavity mode with same field strength.

For experiments without a static external magnetic field, we carried out first measurements with superconducting cavities fabricated out of aluminum yielding quality factors up to $10^5$. They offer much higher $Q$-factors, since dissipation mechanisms are only governed by AC losses in the superconducting material. As a consequence these cavities offer higher sensitivity to emitters in the mode volume, while maintaining the advantages of large and homogeneous coupling rates.


\section{\label{sec:conclusion}Conclusion}

We have presented a novel design for a 3D lumped element resonator for cavity QED experiments. They offer high single spin coupling rates combined with a homogeneous field distribution throughout their mode volume. 
We fabricated the cavity out of copper, which allows Zeeman tuning of the coupled spin system with an external magnetic field. These cavities already yield reasonable high $Q$-values. Even higher $Q$-values, albeit no external magnetic field tunability, are achievable when using superconducting cavities.
With the large coupling rates of this cavity we were able to enter the strong coupling regime of cavity QED using an ensemble of NV centers in diamond. In addition, the achieved homogeneity of the cavity field is a requirement to perform coherent spin manipulation on the whole spin ensemble and implement cavity QED protocols where this becomes important.

\section*{\label{sec:citeref}Acknowledgments}
We would like to thank J\"org Schmiedmayer, William J. Munroe and Michael Trupke for fruitful discussions and Hiroshi Abe from the Japan Agency for Quantum and Radiological Science and Technology for assistance with electron irradiation. The experimental effort has been supported by the TOP grant of TU Wien and the Japan Society for the Promotion of Science KAKENHI (No. 26246001). A.A., T.A. and S.P. acknowledge support by the Austrian Science Fund (FWF) in the framework of the Doctoral School "Building Solids for Function" (Project W1243). We especially thank H. Hartmann, R. Gergen, R. Flasch, A. Linzer and W. Klikovich of the mechanical workshop of the Institute of Atomic and Subatomic Physics for technical assistance.

\clearpage 

\clearpage

\bibliographystyle{mynaturemag}	
\bibliography{citations.bib}	
	

\end{document}